\documentclass[conference]{IEEEtran}

\usepackage{cite}
\usepackage{amsmath,amssymb,amsfonts}
\usepackage{algorithmic}
\usepackage{graphicx}
\usepackage{tikz-layers}
\usepackage{textcomp}
\usepackage{xcolor}
\def\BibTeX{{\rm B\kern-.05em{\sc i\kern-.025em b}\kern-.08em
    T\kern-.1667em\lower.7ex\hbox{E}\kern-.125emX}}
\usepackage{tikz}
\tikzstyle{pinstyle} = [pin edge={to-,thin,black}]
\usetikzlibrary{calc}
\newlength\fheight
\newlength\fwidth
\usepackage{pgfplots}
\pgfplotsset{compat=1.17}
\usepackage{cite}
\usepackage{breakurl}
\usepackage{url}
\usepackage{amsmath,amsfonts,amssymb,amsthm, bm}

    \usepackage{subfig}

\usepackage{pifont}

\usepackage{multirow}
\usepackage{rotating}

\DeclareMathSymbol{\smi}{\mathbin}{AMSa}{"39}

\usepackage{balance}

\usepackage{xcolor}
\usepackage{multirow}
\usepackage{array}
\usepackage{arydshln}
\usepackage{tikz,pgfplots}
\usepackage{tikz-3dplot}
\tikzset{>=latex}
\usepgfplotslibrary{groupplots}
\usetikzlibrary{calc}
\pgfdeclarelayer{background}
\pgfdeclarelayer{foreground}
\pgfsetlayers{background,main,foreground} 
\usetikzlibrary{shapes}
\usetikzlibrary {shapes.multipart}
\usetikzlibrary{positioning}
\usepgfplotslibrary{fillbetween}
\usepackage{tabularx}

\usepackage{caption}
\newcommand\pro[1]{\mbox{#1\%}}

\DeclareMathSymbol{\shortminus}{\mathbin}{AMSa}{"39}
 
\pgfplotsset{compat=1.16,}
\usepackage{makecell}
\usepackage{textcomp}
\usepackage[absolute]{textpos}

\newcommand{\copyrightstatement}{
    \begin{textblock}{15}(0.5,0.3)    
         \noindent
         \centering
         \textblockcolour{white}
         \footnotesize
         \copyright 2024 IEEE. Personal use of this material is permitted. Permission from IEEE must be obtained for all other uses, in any current or future media, including reprinting/republishing this material for advertising or promotional purposes, creating new collective works, for resale or redistribution to servers or lists, or reuse of any copyrighted component of this work in other works.
    \end{textblock}
}

\begin{document}

\copyrightstatement
\title{A Comprehensive Review of Software and Hardware Energy Efficiency of Video Decoders}

\author{\IEEEauthorblockN{Matthias Kr\"anzler, Christian Herglotz, and Andr\'e Kaup}
	\IEEEauthorblockA{Chair of Multimedia Communications and Signal Processing, \\
		Friedrich-Alexander-Universität Erlangen-N\"urnberg (FAU)\\
		\{matthias.kraenzler, christian.herglotz, andre.kaup\}@fau.de}}

\maketitle

\begin{abstract}
  Energy and compression efficiency are two essential parts of modern video decoder implementations that have to be considered. This work comprehensively studies the following six video coding formats regarding compression and decoding energy efficiency: AVC, VP9, HEVC, AV1, VVC, and AVM. We first evaluate the energy demand of reference and optimized software decoder implementations. Furthermore, we consider the influence of the usage of SIMD instructions on those decoder implementations. We find that AV1 is a sweet spot for optimized software decoder implementations with an additional energy demand of \pro{16.55} and bitrate savings of \pro{-43.95} compared to VP9. We furthermore evaluate the hardware decoding energy demand of four video coding formats. Thereby, we show that AV1 has energy demand increases by \pro{117.50} compared to VP9. For HEVC, we found a sweet spot in terms of energy demand with an increase of \pro{6.06} with respect to VP9. Relative to their optimized software counterparts, hardware video decoders reduce the energy consumption to less than \pro{9} compared to software decoders.
\end{abstract}

\IEEEpeerreviewmaketitle

\section{Introduction}
\label{sec:intro}
In recent years, video communication has become increasingly popular. Global mobile communication over the Internet is estimated to increase from 39.5 EB/month in 2019 to 472 EB/month in 2028~\cite{Ericson2023}, a ten fold increase within less than a decade. This rise in data traffic is caused by video content streaming, which made up \pro{71} of the total data traffic in 2023 and is estimated to increase to \pro{80} in 2028. 

Another study showed that video communications caused \pro{1} of global greenhouse gas emissions in 2018~\cite{Efoui-Hess2019}. Therefore, it is essential to consider the energy efficiency of video communications as a globally important sustainability factor. Furthermore, energy efficiency is important from a consumer perspective for mobile devices with a limited battery life. In~\cite{Herglotz19a}, it was shown that video content consumption leads to a significant reduction in terms of battery life.

One of the main goals of the standardization bodies of ISO/ITU or the Alliance for Open Media (AOM) is reducing data traffic in video streaming. Therefore, both organisations study methods and techniques to increase the compression efficiency of state-of-the-art video coders and propose new algorithms with considerable improvements. In 2003, ISO/ITU finalized the specification of the video coding format Advanced Video Coding (AVC)~\cite{Wiegand2003}, which is still broadly used in video communication. After AVC, the following video format was High Efficiency Video Coding (HEVC)~\cite{Sullivan2012}, published in 2013. The most recent video coding format of ISO/ITU is Versatile Video Coding (VVC)~\cite{Bross2021a}, which was standardized in 2020. In each generation of the video coding formats, the target was to reduce the data rate by \pro{50} at equivalent subjective quality.

In 2013, the first version of the royalty-free video coding format VP9~\cite{Mukherjee2013} was published by Google. The successor of VP9 was AOMedia Video 1 (AV1), the first video coding format AOM specified in 2018~\cite{Chen2020}. AV1 is also a royalty-free and open video standard. There are plans to publish the successor of AV1, which is called AOM Video Model (AVM), at the time of this work. 

\begin{table}[!t]
\caption{Overview of the evaluated encoder and decoder SW implementations with the corresponding version.}
\vspace{-0.3cm}
    \label{tab:SoftwareCoDecs}
    \begin{center}
    \begin{tabular}{ l | l | l | l } 
    &  Encoder &  Ref. Decoder &  Optimized Decoder \\
    \hline \hline
    AVC  & x264 (r3065)~\cite{x264} & JM (19.1)~\cite{JM} & FFmpeg (4.4)~\cite{ffmpeg} \\
    \hdashline
    HEVC\! & x265 (3.5.1)~\cite{x265} & HM (16.23)~\cite{HM} & openHEVC (2.0)~\cite{openHEVC}\!\! \\
    \hdashline
    VVC  & VVenC (1.7)~\cite{VVenCPaper} & VTM (19.0)~\cite{VTM}\! & VVdeC (1.6)~\cite{VVdeCPaper} \\
    \hdashline
    VP9  & libvpx (1.10)~\cite{libvpx}\! & libvpx (1.10) & FFmpeg (4.4) \\
    \hdashline
    AV1  & libaom (3.3)~\cite{libaom} & libaom (3.3) & dav1d (1.0)~\cite{dav1d} \\
    \hdashline
    AVM  & avm (3.0)~\cite{avm} & avm (3.0) & / \\
    \end{tabular}
    \end{center}
    \vspace{-0.7cm}
  \end{table}

In the literature, the complexity and compression efficiency of several coding standards was studied before. In~\cite{Laude_2019}, the video codecs HEVC, AVC, VVC, and AV1 were studied regarding encoding time, decoding time, and compression efficiency for software (SW) implementations. In~\cite{Kraenzler2020MMSP}, the energy demand of HEVC and VVC video decoders was studied, and optimization of the energy demand was also proposed. In~\cite{KhernacheBenmoussaBoukhobzaEtAl2021}, it was analyzed to which extent the energy demand of HEVC decoding influences the total device energy depending on SW or hardware (HW) decoding usage. It was found that HEVC SW decoding consumes up to four times more energy than HW decoding. \cite{KhernacheBenmoussaBoukhobzaEtAl2021} looks at the full device power consumption including static power, in our work we are only looking at the dynamic energy consumption. Furthermore, it was shown that the HW decoder consumes up to \pro{30} of the devices power consumption, whereas the SW decoder consumes \pro{50}. Other works studied VVC in detail compared with HEVC~\cite{Mercat_2021} or AV1~\cite{Nguyen_2021}. Another approach in~\cite{Katsenou_2022} studied how the energy demand and the bit rate of AV1, VVC, VP9, and HEVC can be described by a single metric. Therefore, a tradeoff between energy and bitrate is possible across multiple video codecs.

\begin{table*}[!t]

  \caption{Encoder settings for the test conditions random access (RA) and low delay B (LB) according to the AOM CTCs. The used QP values are given in the third column. We omitted the sequence-specific parameters because those have to be set individually for each sequence. In blue, we denote the part of the command only used for RA and in red for LB.}
  \centering
  \begin{tabular}{p{.05\textwidth} | p{.8\textwidth} | p{.057\textwidth}}    
  Codec &  \multicolumn{1}{c|}{Command} & QP \\
  \hline
  x264 & 
  $\shortminus \shortminus$profile=high10  $\shortminus \shortminus$preset=placebo $\shortminus \shortminus$psnr $\shortminus \shortminus$tune=psnr $\shortminus \shortminus$no$\shortminus$scenecut  $\shortminus \shortminus$pass=1 \color{blue}{$\shortminus \shortminus$keyint=65}\color{red}{$\shortminus \shortminus$keyint=infinite $\shortminus \shortminus$min$\shortminus$keyint=$\shortminus$1} & 22 27 32 37 42 47 \\
  \hline
  x265 & $\shortminus \shortminus$profile=main10 $\shortminus$D=10 $\shortminus \shortminus$preset=placebo $\shortminus \shortminus$psnr $\shortminus \shortminus$tune=psnr $\shortminus \shortminus$pools $\shortminus \shortminus$rd=6 $\shortminus \shortminus$rect $\shortminus \shortminus$amp pass=1 $\shortminus \shortminus$frame$\shortminus$threads=1 \color{blue}{$\shortminus \shortminus$keyint=65 $\shortminus \shortminus$min$\shortminus$keyint=65} \color{red}{$\shortminus \shortminus$keyint=65 $\shortminus \shortminus$min$\shortminus$keyint=65} & 22 27 32 37 42 47  \\
  \hline
  vvenc &$\shortminus$rs 1 $\shortminus \shortminus$qpa 1 \color{blue}{$\shortminus$c cfg/randomaccess\_slower.cfg $\shortminus \shortminus$IntraPeriod=65} \color{red}{$\shortminus$c cfg/experimental/lowdelay\_slower.cfg\!} & 22 27 32 37 42 47  \\
  \hline
  libvpx & $\shortminus \shortminus$cpu$\shortminus$used=1 $\shortminus \shortminus$passes=1  $\shortminus \shortminus$auto$\shortminus$alt$\shortminus$ref=1 $\shortminus \shortminus$min$\shortminus$gf$\shortminus$interval=16 $\shortminus \shortminus$max$\shortminus$gf$\shortminus$interval=16   $\shortminus \shortminus$enable$\shortminus$tpl=0 $\shortminus \shortminus$end$\shortminus$usage=q  $\shortminus \shortminus$profile=2 $\shortminus$b 10 \color{blue}{$\shortminus \shortminus$lag$\shortminus$in$\shortminus$frames=19 $\shortminus \shortminus$kf$\shortminus$min$\shortminus$dist=65 $\shortminus \shortminus$kf$\shortminus$max$\shortminus$dist=65} \color{red}{$\shortminus \shortminus$lag$\shortminus$in$\shortminus$frames=0 $\shortminus \shortminus$kf$\shortminus$max$\shortminus$dist=9999 $\shortminus \shortminus$kf$\shortminus$min$\shortminus$dist=9999} & 23 31 39 47 55 63  \\
  \hline
  libaom \& avm &  $\shortminus \shortminus$cpu$\shortminus$used=0 $\shortminus \shortminus$passes=1 $\shortminus$b 10 $\shortminus \shortminus$obu $\shortminus \shortminus$min$\shortminus$gf$\shortminus$interval=16 $\shortminus \shortminus$max$\shortminus$gf$\shortminus$interval=16  $\shortminus \shortminus$gf$\shortminus$min$\shortminus$pyr$\shortminus$height=4 $\shortminus \shortminus$gf$\shortminus$max$\shortminus$pyr$\shortminus$height=4  $\shortminus \shortminus$use$\shortminus$fixed$\shortminus$qp$\shortminus$offsets=1 $\shortminus \shortminus$deltaq$\shortminus$mode=0 $\shortminus \shortminus$enable$\shortminus$tpl$\shortminus$model=0 $\shortminus \shortminus$end$\shortminus$usage=q $\shortminus \shortminus$enable$\shortminus$keyframe$\shortminus$filtering=0  \color{blue}{ $\shortminus\shortminus$lag$\shortminus$in$\shortminus$frames=19 $\shortminus\shortminus$auto$\shortminus$alt$\shortminus$ref=1  $\shortminus\shortminus$kf$\shortminus$min$\shortminus$dist=65 
  $\shortminus\shortminus$kf$\shortminus$max$\shortminus$dist=65} \color{red}{$\shortminus\shortminus$lag$\shortminus$in$\shortminus$frames=0  $\shortminus\shortminus$kf$\shortminus$min$\shortminus$dist=9999 $\shortminus\shortminus$kf$\shortminus$max$\shortminus$dist=9999 }& 23 31 39 47 55 63    \\
\end{tabular}
  \label{tab:Commands}
  \vspace{-0.5cm}
\end{table*}

In this work, we will thoroughly study the energy and compression efficiency of six video codecs and study 11 SW decoder implementations with and without SIMD instructions enabled. Furthermore, we extend the analysis by comparing the energy efficiency of reference and optimized SW video decoders. Additionally, we will look at the energy efficiency of HW decoding. Furthermore, we will compare the energy demand of SW decoders in relation to HW decoding.

Section~\ref{sec:Setup} will introduce the used setup and the applied metrics. In Section~\ref{sec:Analysis}, we will analyze the compression of each video coding format and energy efficiency for SW and HW decoding. Furthermore, we will compare the energy efficiency of HW and SW decoding. Finally, in Section~\ref{sec:Conclusion}, we will summarize the results and give an outlook to future research.

\section{Setup and Metrics}
\label{sec:Setup}

\subsection{Video Encoder and Decoder Setup}
\label{subsec:VideoCodecs}
\noindent In this work, we evaluate six video coding standards. In Table~\ref{tab:SoftwareCoDecs}, we show which encoders and decoders are used for generating and measuring the bitstreams for each video codec. We selected one decoder used during the standardization process as a reference implementation. Furthermore, we evaluate a second optimized decoder used for practical applications. In the table, we also provide the version tags that were used. We also evaluate the influence of the Single Instruction Multiple Data (SIMD) instruction set on the energy efficiency of decoders.

In Table~\ref{tab:Commands}, we show the commands used for each encoder in this work. With the encoders in Table~\ref{tab:SoftwareCoDecs}, we encode the video sequences that are suggested by the common test conditions (CTCs) of AOM~\cite{CWGB005oV1}. From these CTCs, we take the video sequences of class A1 with 4K resolution, class A2 with full HD resolution, class A3 with HD resolution, and class B with full HD resolution. We encode with the random access (RA) test conditions and low delay B (LB). The command parts in blue are only used for RA, the red ones only for LB. For the libaom and the AVM encoder, we use the same commands to encode the bitstreams. The quantization parameter (QP) values we used are given in the third column.

\subsection{Measurement Setups}
\label{subsec:Measurement}
\noindent To evaluate energy efficiency, we conduct energy measurements described as follows. We first measure the energy demand during the decoding process and then the energy demand during idle. Next, the idle energy demand is subtracted from the first measurement. More details on the measurement method can be found in~\cite{HerglotzSpringerReichenbachEtAl2018}. For statistical correctness, we check the measurement series for one bitstream with a confidence interval test with a maximum deviation of less than \pro{2} from the actual values with a probability of \pro{99} as described in~\cite{HerglotzSpringerReichenbachEtAl2018}. For a fair comparison of each decoder, we limit the decoders to single-thread execution.

For SW decoding measurements, we use a desktop PC with an Intel i7-8700 CPU and CentOS as an operating system (OS). The CPU incorporates an internal power meter with Running Average Power Limit (RAPL)~\cite{DavidGorbatovHanebutteEtAl2010} that can be directly accessed for energy measurements by the OS. Previous experiments showed that the measurements of RAPL align accurately with an external power meter~\cite{JVET-P0084}.

For the HW decoder measurements, we use the Rock 5 Model B single board computer (SBC), which provides an octa-core ARM processor and a video processing unit (VPU) that supports the decoding of AVC, VP9, HEVC, and AV1 in HW. As OS, we use Ubuntu, and the HW decoder is accessed over FFmpeg. The SBC is connected over the main power supply jack to derive the energy demand with an external power meter, the LMG 611, by ZES Zimmer.

\subsection{Evaluation Metrics}
\label{subsec:Metrics}

\noindent Apart from the complexity of SW implementations, the compression performance of the codec is important for standardization. A commonly used metric to describe the compression gains at a given objective visual quality level is the Bj{\o}ntegaard Delta (BD) metric~\cite{VCEG-M33}. The BD metric is computed with the framework proposed in~\cite{Herglotz2023,Herglotz2023a}, which proposes an improved interpolation method for the BD rate (BDR) and BD decoding energy (BDDE), where BDDE describes the decoding energy savings or increases for the same visual quality level. In this work, negative BD values correspond to savings in terms of bit rate or energy demand. Therefore, we use the YUV-PSNR to measure the visual quality according to~\cite{WorkingPractices}. As an anchor for the BD metric calculations, we will use VP9 throughout this paper.

\section{Energy and Compression Analysis}
\label{sec:Analysis}
\noindent In the following, we will first evaluate the compression efficiency of each video codec in Section~\ref{subsec:CompressionEfficiency}, then the SW energy demand in Section~\ref{subsec:SoftwareEfficiency}. Thereafter, we assess the HW energy demand of each video decoder in Section~\ref{subsec:HardwareEfficiency}. Finally, we compare the energy demand of SW decoding with HW decoding in Section~\ref{subsec:GlobalEval}.

\begin{table}[!t]

  \caption{ Evaluation of the compression and energy efficiency for all video codecs in relation to VP9. For the evaluation of BDR, we compare the bitstreams of each codec with the encoded bistreams of VP9. For the BDDE values, we utilize the decoding energy demand of libvpx as an anchor for the reference decoders. For the optimized decoders, we use the decoding of VP9 with FFmpeg as an anchor. Furthermore, we distinguish between decoders with SIMD instructions enabled (S.on) and SIMD disabled (S.off). In the final column, the BDDE results of the hardware decoders are shown, which are compared with VP9 hardware decoding. If a decoder was unavailable, we marked the corresponding values as N/A. All BD-values are given in \%.}
  \vspace{-0.3cm}
  \label{tab:ComplexityAnalysis}
  \begin{center}
  \begin{tabular}{l || r |  r :  r |  r :  r||  r} 
 
    & & \multicolumn{2}{c | }{SW Reference} & \multicolumn{2}{c || }{SW Optimized} & \!\!Hardware\!\! \\
    & & \multicolumn{1}{c:}{S.on} & \multicolumn{1}{c|}{S.off} & \multicolumn{1}{c:}{S.on} & \multicolumn{1}{c||}{S.off} & \\
   Codec & \multicolumn{1}{c|}{BDR} & \multicolumn{1}{c:}{\!\!BDDE\!\!}
    & \multicolumn{1}{c|}{\!\!BDDE\!\!}& \multicolumn{1}{c:}{\!\!BDDE\!\!}
    & \multicolumn{1}{c||}{\!\!BDDE\!\!}& \multicolumn{1}{c}{\!\!BDDE\!\!}\\ 
  \hline \hline 
  & \multicolumn{6}{c}{Random access} \\
   \hline 
AVC & 27.09 &  N/A &  -6.34 &  -27.56 &  -33.05 & -24.47 \\ 
VP9 & 0.00 &  0.00 &  0.00 &  0.00 &  0.00 & 0.00 \\ 
HEVC & -17.47 &  N/A &  34.93 &  18.29 &  6.07 & 6.06 \\ 
AV1 & -43.95 &  95.31 &  212.69 &  16.55 &  82.29 & 117.50 \\ 
VVC & -52.07 &  500.31 &  232.15 &  198.29 &  241.18 & N/A \\ 
AVM & -48.38 &  247.15 &  339.16 &  N/A &  N/A & N/A \\ 
  \hline \hline 
  & \multicolumn{6}{c}{Low delay B} \\
  \hline     
  AVC & 21.71 &  N/A &  -7.45 &  -30.95 &  -35.69 & -26.74  \\ 
  VP9 & 0.00 &  0.00 &  0.00 &  0.00 &  0.00 & 0.00 \\ 
  HEVC & -16.39 &  N/A &  37.95 &  20.31 &  6.53 & 6.69 \\ 
  AV1 & -28.90 &  124.71 &  239.17 &  30.78 &  101.72 & 109.06 \\ 
  VVC & -33.45 &  515.79 &  215.43 &  197.73 &  201.15 & N/A \\ 
  AVM & -33.40 &  225.38 &  274.95 &  N/A &  N/A & N/A \\ 
  \end{tabular}
  \end{center}

  \vspace{-0.7cm}
\end{table}

\subsection{Compression Efficiency Evaluation}
\label{subsec:CompressionEfficiency}
\noindent Table~\ref{tab:ComplexityAnalysis} shows an overview of all results for the comparison the video codecs in terms of compression and energy efficiency. First, we will evaluate the compression efficiency in terms of BDR, which is shown in the second column. For the calculation of BDR, we use the encoded bitstreams of VP9 as an anchor. From Table~\ref{tab:ComplexityAnalysis}, we can determine that AVC has the lowest compression efficiency (BDR: \pro{27.09}) compared to VP9 for RA test condition. With HEVC as a successor of AVC, the compression efficiency was improved significantly (BDR: \pro{-17.47}). With AV1, there is a significant improvement of over \pro{-40} in BDR over VP9. For VVC and AVM, we determined half the required data rate for the same PSNR compared to VP9 with \pro{-52.07} for VVC and with \pro{-48.38} for AVM. 

For the LB test condition, we determine that the improvements in terms of BDR are smaller compared to the previous values of RA. For HEVC, the BDR is similar with a BDR of \pro{-16.39}. For VVC and AVM, the improvements reach a BDR of up to \pro{-33}  compared to VP9.

\subsection{Software Decoder Evaluation}
\label{subsec:SoftwareEfficiency}

In the following, we compare the decoding energy demand of several video decoders with the corresponding implementations of VP9, which is libvpx for the reference implementations and FFmpeg for the optimized decoders. Also, we distinguish between SIMD-enabled and SIMD-disabled decoders in Table~\ref{tab:ComplexityAnalysis}. HEVC (HM) and AVC (JM) reference decoder implementations do not use SIMD optimizations. Therefore, we set the corresponding values to not available (N/A).

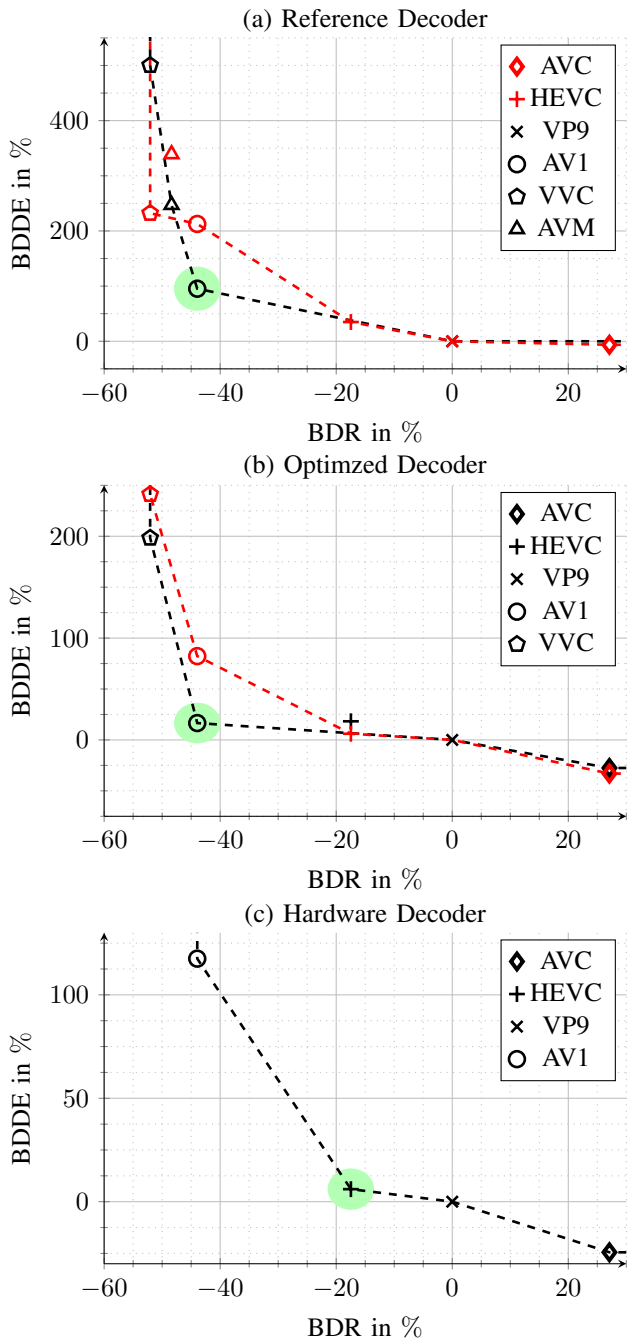
\begin{figure}[!t]

  \begin{center}
  \begin{tikzpicture}
  \begin{groupplot}[
       group style={group name=my plots,group size= 1 by 3,vertical sep =1.55cm},
        title style={yshift=-0.25cm},
       height = 10cm,
       width = \textwidth ,set layers,cell picture=true
      ]
      
  \nextgroupplot[
  width=0.47\textwidth,
  height = 0.33\textwidth,
  xlabel={BDR in $\%$ },
  title = {(a) Reference Decoder},
  ylabel={BDDE in $\%$},
  xmin=-60, xmax=30,
  ymin=-50, ymax=550,
  axis lines = left,
  ymajorgrids=true,
  yminorgrids=true,
  xmajorgrids=true,
  xminorgrids=true,
  minor tick num=3,
  minor grid style=dotted,
  ]

  \addplot[only marks,draw=black,
  color=red,mark=diamond,
  line width=1.5pt,mark size =3pt,]
  coordinates {(27.09,-6.34)};
  \addlegendentry{AVC}

  \addplot[only marks,draw=black,
  color=red,mark=+,
  line width=1pt,mark size =3pt,]
  coordinates {(-17.47,34.93)};
  \addlegendentry{HEVC}

  \addplot[only marks,draw=black,
  color=black,mark=x,
  line width=1pt,mark size =3pt,]
  coordinates {(0,0)};
  \addlegendentry{VP9}

  \addplot[only marks,draw=black,
  color=black,mark=o,
  line width=1pt,mark size =3pt,]
  coordinates {(-43.95,95.31)};
  \addlegendentry{AV1}
  \fill[color=green!30, fill=green!30] (-43.95,95.31) ellipse (4 and 40);
  \addplot[only marks,draw=black,
  color=black,mark=pentagon,
  line width=1pt,mark size =3pt,]
  coordinates {(-52.07,500.31)};
  \addlegendentry{VVC}    

  \addplot[only marks,draw=black,
  color=black,mark=triangle,
  line width=1pt,mark size =3pt,]
  coordinates {(-48.38,247.15)};
  \addlegendentry{AVM}


  \addplot[only marks,draw=black,
  color=red,mark=x,
  line width=1pt,mark size =3pt,]
  coordinates {(0,0)};

  \addplot[only marks,draw=black,
  color=red,mark=o,
  line width=1pt,mark size =3pt,]
  coordinates {(-43.95,212.69)};

  \addplot[only marks,draw=black,
  color=red,mark=pentagon,
  line width=1pt,mark size =3pt,]
  coordinates {(-52.07,232.15)};

  \addplot[only marks,draw=black,
  color=red,mark=triangle,
  line width=1pt,mark size =3pt,]
  coordinates {(-48.38,339.16)  };

  \addplot [ 
          color=black,
          mark=none,
          mark size=2pt,
          line width=1pt,dashed
  ] coordinates { (40,0) (0,0) (-43.95,95.31) (-48.38,247.15)       
                (-52.07,500.31)  (-52.07,600)				
          };
  
  \addplot [ 
      color=red,
      mark=none,
      mark size=2pt,
      line width=1pt,dashed
  ] coordinates { (40,-6.34) (27.09,-6.34) (0,0) (-17.47,34.93)
  (-43.95,212.69) (-52.07,232.15) (-52.07,600)				
          };
  \nextgroupplot[
      width=0.47\textwidth,
      height = 0.33\textwidth,
      xlabel={BDR in $\%$ },
      title = {(b) Optimzed Decoder},
      ylabel={BDDE in $\%$},
      xmin=-60, xmax=30,
      ymin=-75, ymax=250,
      axis lines = left,
      ymajorgrids=true,
      yminorgrids=true,
      xmajorgrids=true,
      xminorgrids=true,
      minor tick num=3,
      minor grid style=dotted,
      ]
      \addplot[only marks,draw=black,
  color=black,mark=diamond,
  line width=1.5pt,mark size =3pt,]
  coordinates {(27.09,-27.56)};
  \addlegendentry{AVC}

  \addplot[only marks,draw=black,
  color=black,mark=+,
  line width=1pt,mark size =3pt,]
  coordinates {(-17.47,18.29)};
  \addlegendentry{HEVC}

  \addplot[only marks,draw=black,
  color=black,mark=x,
  line width=1pt,mark size =3pt,]
  coordinates {(0,0)};
  \addlegendentry{VP9}

  \addplot[only marks,draw=black,
  color=black,mark=o,
  line width=1pt,mark size =3pt,]
  coordinates {(-43.95,16.55)};
  \addlegendentry{AV1}
  \fill[color=green!30, fill=green!30] (-43.95,16.55) ellipse (4 and 20);

  \addplot[only marks,draw=black,
  color=black,mark=pentagon,
  line width=1pt,mark size =3pt,]
  coordinates {(-52.07,198.29)};
  \addlegendentry{VVC}    

  \addplot[only marks,draw=black,
  color=red,mark=diamond,
  line width=1.5pt,mark size =3pt,]
  coordinates {(27.09,-33.05)};

  \addplot[only marks,draw=black,
  color=red,mark=+,
  line width=1pt,mark size =3pt,]
  coordinates {(-17.47,6.07)};

  \addplot[only marks,draw=black,
  color=red,mark=o,
  line width=1pt,mark size =3pt,]
  coordinates {(-43.95,82.29)};

  \addplot[only marks,draw=black,
  color=red,mark=pentagon,
  line width=1pt,mark size =3pt,]
  coordinates {(-52.07,241.18)};

  \addplot [ 
  color=black,
  mark=none,
  mark size=2pt,
  line width=1pt,dashed
] coordinates { (30,-27.56) (27.09,-27.56) (0,0) (-43.95,16.55)       
        (-52.07,198.29)  (-52.07,300)				
  };

\addplot [ 
color=red,
mark=none,
mark size=2pt,
line width=1pt,dashed
] coordinates { (40,-33.05) (27.09,-33.05) (0,0) (-17.47,6.07)
(-43.95,82.29) (-52.07,241.18) (-52.07,300)};
\vspace{-1cm}
      \nextgroupplot[
          width=0.47\textwidth,
          height = 0.33\textwidth,
          xlabel={BDR in $\%$ },
          title = {(c) Hardware Decoder},
          ylabel={BDDE in $\%$},
          xmin=-60, xmax=30,
          ymin=-30, ymax=130,
          axis lines = left,
          ymajorgrids=true,
          yminorgrids=true,
          xmajorgrids=true,
          xminorgrids=true,
          minor tick num=3,
          minor grid style=dotted,
          ]
          
      \addplot[only marks,draw=black,
      color=black,mark=diamond,
      line width=1.5pt,mark size =3pt,]
      coordinates {(27.09,-24.47)};
      \addlegendentry{AVC}
  
      \addplot[only marks,draw=black,
      color=black,mark=+,
      line width=1pt,mark size =3pt,]
      coordinates {(-17.47,6.06)};
      \addlegendentry{HEVC}
  
      \fill[color=green!30, fill=green!30] (-17.47,6.06) ellipse (4 and 10);

      \addplot[only marks,draw=black,
      color=black,mark=x,
      line width=1pt,mark size =3pt,]
      coordinates {(0,0)};
      \addlegendentry{VP9}
  
      \addplot[only marks,draw=black,
      color=black,mark=o,
      line width=1pt,mark size =3pt,]
      coordinates {(-43.95,117.50)};
      \addlegendentry{AV1}

      \addplot [ 
      color=black,
      mark=none,
      mark size=2pt,
      line width=1pt,dashed
  ] coordinates { (30,-24.47) (27.09,-24.47) (0,0) (-17.47,6.06) (-43.95,117.50)         (-43.95,130)				
      };
  \end{groupplot}

  \end{tikzpicture}
  \end{center}
  \vspace{-0.4cm}
  \caption{Evaluation of the compression and energy efficiency in terms of BDR and BDDE for RA test condition. In (a), we evaluate the results of the reference decoders, in (b) for the optimized decoders, and (c) for the HW decoder. The vertical axis denotes the BDDE values compared to VP9, and the horizontal axis shows the BDR. The red markers show the results for the decoder with SIMD disabled and the black markers with SIMD enabled. For each option, we show a Pareto front that shows the optimal energy efficiency for a given compression efficiency. The green are corresponds to a sweet spot that represents a good tradeoff between compression and energy efficiency.}
  \vspace{-0.3cm}

  \label{fig:Analysis}
  \end{figure}

Furthermore, we illustrate the results in Figure~\ref{fig:Analysis}. In \mbox{Figure \ref{fig:Analysis}(a)}, we show the evaluation for the reference decoders, in \mbox{Figure \ref{fig:Analysis}(b)}, for the optimized decoder implementations, and in \mbox{Figure \ref{fig:Analysis}(c)}, for the HW decoders. On the vertical axis of each graph, we show the BDDE values, and on the horizontal axis, we show the BDR values. Furthermore, we distinguish between decoders with SIMD enabled (black markers) and SIMD disabled (red markers). The Pareto front of each option provides the most energy-efficient video codec for a given BDR value.

For reference decoders with SIMD disabled, we determine that the energy efficiency of AVC is slightly improved over VP9 (BDDE: \pro{-6.34}), and for HEVC it is increased by \pro{34.93}. For the optimized decoder implementations, we determine that AVC improves the energy demand for SIMD enabled (BDDE: \pro{-27.56}) and disabled decoding (BDDE: \pro{-33.05}). For HEVC, the increase of the decoding energy demand is significantly lower than for the reference decoder implementations. In Figure~\ref{fig:Analysis}, we determine that HEVC decoding (plus markers) is not part of the Pareto front if SIMD instructions are enabled. For AVC (diamond marker), we observe that the SW decoders are always part of the Pareto front, representing the lowest energy demand.

For AV1, the energy demand of the reference decoder implementation is almost doubled (BDDE: \pro{95.31}). Nevertheless, for the optimized implementation of AV1 with SIMD enabled, the energy demand is increased by a similar level as for HEVC with a BDDE of \pro{16.55}. For reference and optimized decoders in Figure~\ref{fig:Analysis}, we observe that AV1 (circle markers) is a sweet spot, which is highlighted in green in the graph. We find that it is more reasonable to use AV1 over HEVC if energy and compression efficiency is targeted for SW decoders. However, if SIMD is disabled, the energy demand is increased to \pro{82.29}, which is significantly higher than the corresponding value for HEVC (\pro{6.07}).

For VVC and AVM, we observe that the energy demand of the decoder implementations is increased by multiple times compared to VP9. With an additional energy demand of \pro{198.29} for the optimized implementation, the energy demand of VVdeC is three times the energy demand of FFmpeg for VP9. This shows that coding tools introduced to VVC are very complex in SW, and we observe that the energy demand couldn't be reduced to the same level as for VP9 by optimizing the implementation. In Figure~\ref{fig:Analysis}, we observe a steep increase of the energy demand for lower BDR values.

\subsection{Hardware Decoder Evaluation}
\label{subsec:HardwareEfficiency}
\noindent To evaluate the HW decoder energy efficiency, we show the BDDE values, which are compared with VP9 hardware decoding, in the final column of Table~\ref{tab:ComplexityAnalysis}. We find that the BD values of the test conditions RA and LB are similar. For AVC, the energy demand is reduced on average by \pro{-24.47} for RA and by \pro{-26.74} for LB, accordingly. Therefore, we see that the energy efficiency of AVC is higher than that of VP9, as is the case for SW decoding with optimized decoders. For HEVC, the decoding demand is slightly higher than for VP9 with a BDDE of \pro{6.06} for RA. This value is also close to the values of the optimized decoder (BDDE: \pro{18.29}).

However, despite the close alignments of AVC and HEVC for the Rock board, we determine that the energy demand of AV1 is significantly increased compared to VP9 with a BDDE of \pro{117.50}. We observed that some sequences failed to be decoded by the HW decoder\footnote{The following sequences are sorted out for the HW BD metrics: MINECRAFT, Mission Control, Wikipedia, Vertical Bees, Vertical Carnaby, and WalkingInStreet.}. The BDDE values of HW decoding have similar values to the optimized SW decoders with SIMD off, which have a BDDE of \pro{82.29} for AV1. Based on those results we observe that the optimized decoder with SIMD off for VVC has a BDDE \pro{241.18}. For future HW implementations of VVC, this would imply that there will be a significantly increased energy demand over VP9, which could be even higher than AV1.

\subsection{Software vs. Hardware Decoder Energy Evaluation}
\label{subsec:GlobalEval}
\noindent So far, the energy efficiency evaluation was limited to comparing the same decoder type across each video codec. Next, we will show the energy improvements that can be reached by changing from SW to HW decoding. Therefore, we calculate the relative energy demand $\Delta e$ as follows,
\begin{equation}
    \Delta e = \frac{1}{M} \sum_{i=1}^{M} \frac{E_{\textrm{dec,HW},i}}{E_{\textrm{dec,SW},i}},
    \label{eq:MAPE}
\end{equation}
where $i$ is the bitstream index from all $M$ bitstreams in the data set, $E_{\textrm{dec,HW},i}$ is the hardware decoder demand, and the software decoder demand $E_{\textrm{dec,SW},i}$ of the bitstream $i$. The results for this evaluation are shown in Table~\ref{tab:GlobalComplexityAnalysis}. 

For the reference decoders of VP9 and AV1 with SIMD enabled, we observe that the HW decoder energy demand is reduced to $\Delta e=$ \pro{4.51}, which is 22 times less than for SW decoding. If SIMD instructions are disabled for reference decoders, the reduction from SW to HW has an even higher $\Delta e$ of approximately \pro{1}, which is a reduction by a factor of 100.

\begin{table}[!t]
  \caption{Relative energy demand if the corresponding HW decoder is used instead of the SW decoder. We marked the corresponding values as N/A if a SW decoder was unavailable.}
  \vspace{-0.3cm}
  \label{tab:GlobalComplexityAnalysis}
  \begin{center}
  \begin{tabular}{l ||  r :  r |  r :  r} 
 
    & \multicolumn{2}{c | }{ Reference} & \multicolumn{2}{c  }{Optimized}  \\
    Codec & \multicolumn{1}{c:}{SIMD On} & \multicolumn{1}{c|}{SIMD Off} & \multicolumn{1}{c:}{SIMD On} & \multicolumn{1}{c}{SIMD Off}  \\
  \hline \hline 
  & \multicolumn{4}{c}{Random access} \\
   \hline 
   AVC &  N/A &  \pro{1.09} &  \pro{5.22} &  \pro{2.08}   \\ 
   VP9 &  \pro{3.77} &  \pro{1.27} &  \pro{4.56} &  \pro{1.64}  \\ 
   HEVC &  N/A &  \pro{1.04} &  \pro{4.44} & \pro{1.74}   \\ 
   AV1 &  \pro{4.51} &  \pro{1.02} &  \pro{8.86} &  \pro{2.24}   \\ 
  \hline \hline 
  & \multicolumn{4}{c}{Low delay B} \\
  \hline 
  AVC &  N/A &  \pro{1.09} &  \pro{5.39} &  \pro{2.15}  \\ 
  VP9 &  \pro{3.87} &  \pro{1.31} &  \pro{4.70} &  \pro{1.69}   \\ 
  HEVC &  N/A &  \pro{1.05} &  \pro{4.48} &  \pro{1.78}   \\ 
  AV1 &  \pro{3.44} &  \pro{0.89} &  \pro{7.07} &  \pro{1.91}   \\ 
  \end{tabular}
  \end{center}

  \vspace{-0.7cm}
\end{table}

For the optimized decoders with SIMD enabled, we determine that HEVC has the highest energy demand improvement from SW to HW with $\Delta e$ of \pro{4.44}, and for AV1, $\Delta e$ is \pro{8.86} from HW to SW. We observe that the improvement factor of HEVC and VP9 is around 22, and for AV1 it is roughly 11.

\section{Conclusion}
\label{sec:Conclusion}
\noindent This work evaluated six video codecs regarding compression and energy efficiency for SW and HW decoding. Thereby, we provided insights into the tradeoffs between compression and energy consumption. The results show that newer codecs such as AV1, VVC, and AVM have significantly improved compression efficiency over older codecs such as AVC and VP9. However, the relative energy efficiency of SW decoder implementations significantly depends on integrating SIMD instructions. If those are available, we determined that AV1 is a sweet spot for a significant improvement of the compression efficiency with a BDR of \pro{-43.95} compared to VP9 and a low increase in terms of energy demand with a BDDE of \pro{95.31} for reference decoders and \pro{16.55} for optimized decoder compared to the corresponding VP9 decoders. For the HW decoders, we could observe such a sweet spot for HEVC with a BDDE of \pro{6.06} and found a significant increase in terms of energy demand for AV1 (BDDE: \pro{117.50}). We also showed that the dynamic energy demand is reduced to less than \pro{9} by switching from SW to HW decoding. In the future, we want to study further the influence of presets on the energy efficiency of video decoders.

\bibliographystyle{IEEEtran}

\begin{thebibliography}{10}
  \providecommand{\url}[1]{#1}
  \csname url@samestyle\endcsname
  \providecommand{\newblock}{\relax}
  \providecommand{\bibinfo}[2]{#2}
  \providecommand{\BIBentrySTDinterwordspacing}{\spaceskip=0pt\relax}
  \providecommand{\BIBentryALTinterwordstretchfactor}{4}
  \providecommand{\BIBentryALTinterwordspacing}{\spaceskip=\fontdimen2\font plus
  \BIBentryALTinterwordstretchfactor\fontdimen3\font minus
    \fontdimen4\font\relax}
  \providecommand{\BIBforeignlanguage}[2]{{%
  \expandafter\ifx\csname l@#1\endcsname\relax
  \else
  \language=\csname l@#1\endcsname
  \fi
  #2}}
  \providecommand{\BIBdecl}{\relax}
  \BIBdecl
  
  \bibitem{Ericson2023}
  \BIBentryALTinterwordspacing
  {Ericson}. (2023, Jun.) Ericson mobility report. [Online]. Available:
    \url{https://www.ericsson.com/49dd9d/assets/local/reports-papers/mobility-report/documents/2023/ericsson-mobility-report-june-2023.pdf}
  \BIBentrySTDinterwordspacing
  
  \bibitem{Efoui-Hess2019}
  \BIBentryALTinterwordspacing
  M.~Efoui-Hess. (2019, Jul.) Climate crisis: The unsustainable use of online
    video. {T}he practical case for digital sobriety. [Online]. Available:
    \url{https://theshiftproject.org/en/article/unsustainable-use-online-video/}
  \BIBentrySTDinterwordspacing
  
  \bibitem{Herglotz19a}
  C.~Herglotz, S.~Coulombe, S.~Vakili, and A.~Kaup, ``Power modeling for virtual
    reality video playback applications,'' in \emph{Proc. IEEE International
    Symposium on Consumer Technology (ISCT)}, Ancona, Italy, Jun. 2019.
  
  \bibitem{Wiegand2003}
  T.~Wiegand, G.~Sullivan, G.~Bjontegaard, and A.~Luthra, ``Overview of the
    h.264/{AVC} video coding standard,'' \emph{{IEEE} Transactions on Circuits
    and Systems for Video Technology}, vol.~13, no.~7, pp. 560--576, jul 2003.
  
  \bibitem{Sullivan2012}
  G.~J. Sullivan, J.-R. Ohm, W.-J. Han, and T.~Wiegand, ``Overview of the high
    efficiency video coding ({HEVC}) standard,'' \emph{IEEE Transactions on
    Circuits and Systems for Video Technology}, vol.~22, no.~12, pp. 1649--1668,
    Dec. 2012.
  
  \bibitem{Bross2021a}
  B.~Bross, Y.-K. Wang, Y.~Ye, S.~Liu, J.~Chen, G.~J. Sullivan, and J.-R. Ohm,
    ``Overview of the {Versatile Video Coding (VVC)} standard and its
    applications,'' \emph{IEEE Transactions on Circuits and Systems for Video
    Technology}, vol.~31, no.~10, pp. 3736--3764, Oct. 2021.
  
  \bibitem{Mukherjee2013}
  D.~Mukherjee, J.~Bankoski, A.~Grange, J.~Han, J.~Koleszar, P.~Wilkins, Y.~Xu,
    and R.~Bultje, ``The latest open-source video codec {VP}9 - an overview and
    preliminary results,'' in \emph{2013 Picture Coding Symposium ({PCS})}, Dec.
    2013.
  
  \bibitem{Chen2020}
  Y.~Chen, D.~Mukherjee, J.~Han, A.~Grange, Y.~Xu, S.~Parker, C.~Chen, H.~Su,
    U.~Joshi, C.-H. Chiang, Y.~Wang, P.~Wilkins, J.~Bankoski, L.~Trudeau,
    N.~Egge, J.-M. Valin, T.~Davies, S.~Midtskogen, A.~Norkin, P.~de~Rivaz, and
    Z.~Liu, ``An overview of coding tools in {AV}1: the first video codec from
    the alliance for open media,'' \emph{{APSIPA} Transactions on Signal and
    Information Processing}, vol.~9, 2020.
  
  \bibitem{x264}
  \BIBentryALTinterwordspacing
  Videolan. {x264 Encoder}. Accessed 2021-09. [Online]. Available:
    \url{https://code.videolan.org/videolan/x264.git}
  \BIBentrySTDinterwordspacing
  
  \bibitem{JM}
  \BIBentryALTinterwordspacing
  {HHI Fraunhofer}. {JM Decoder}. Accessed 2021-09. [Online]. Available:
    \url{https://vcgit.hhi.fraunhofer.de/jvet/JM}
  \BIBentrySTDinterwordspacing
  
  \bibitem{ffmpeg}
  \BIBentryALTinterwordspacing
  {Fast Forwards MPEG (FFmpeg)}. Accessed 2018-11-14. [Online]. Available:
    \url{http://ffmpeg.org/}
  \BIBentrySTDinterwordspacing
  
  \bibitem{x265}
  \BIBentryALTinterwordspacing
  Videolan. {x265 Encoder}. Accessed 2021-09. [Online]. Available:
    \url{http://hg.videolan.org/x265}
  \BIBentrySTDinterwordspacing
  
  \bibitem{HM}
  \BIBentryALTinterwordspacing
  {HHI Fraunhofer}. {HM Decoder}. Accessed 2021-09. [Online]. Available:
    \url{https://vcgit.hhi.fraunhofer.de/jvet/HM}
  \BIBentrySTDinterwordspacing
  
  \bibitem{openHEVC}
  \BIBentryALTinterwordspacing
  {openHEVC}. Accessed 2021-02-25. [Online]. Available:
    \url{https://github.com/OpenHEVC/openHEVC}
  \BIBentrySTDinterwordspacing
  
  \bibitem{VVenCPaper}
  A.~Wieckowski, J.~Brandenburg, T.~Hinz, C.~Bartnik, V.~George, G.~Hege,
    C.~Helmrich, A.~Henkel, C.~Lehmann, C.~Stoffers, I.~Zupancic, B.~Bross, and
    D.~Marpe, ``{VVenC}: An open and optimized {VVC} encoder implementation,'' in
    \emph{Proc. IEEE International Conference on Multimedia Expo Workshops
    (ICMEW)}, pp. 1--2.
  
  \bibitem{VTM}
  \BIBentryALTinterwordspacing
  {Joint Video Exploration Team ({JVET})}. {VVC test model reference software}.
    [Online]. Available:
    \url{https://vcgit.hhi.fraunhofer.de/jvet/VVCSoftware_VTM/}
  \BIBentrySTDinterwordspacing
  
  \bibitem{VVdeCPaper}
  A.~Wieckowski, G.~Hege, C.~Bartnik, C.~Lehmann, C.~Stoffers, B.~Bross, and
    D.~Marpe, ``Towards a live software decoder implementation for the upcoming
    {Versatile Video Coding (VVC)} codec,'' in \emph{Proc. IEEE International
    Conference on Image Processing (ICIP)}, pp. 3124--3128.
  
  \bibitem{libvpx}
  \BIBentryALTinterwordspacing
  Google. {libvpx Codec}. Accessed 2021-10. [Online]. Available:
    \url{https://chromium.googlesource.com/webm/libvpx/}
  \BIBentrySTDinterwordspacing
  
  \bibitem{libaom}
  \BIBentryALTinterwordspacing
  A.~for Open~Media. {libaom Codec}. Accessed 2022-03. [Online]. Available:
    \url{https://aomedia.googlesource.com/aom/}
  \BIBentrySTDinterwordspacing
  
  \bibitem{dav1d}
  \BIBentryALTinterwordspacing
  DAV1D. {DAV1D Software}. Accessed 2022-03. [Online]. Available:
    \url{https://code.videolan.org/videolan/dav1d}
  \BIBentrySTDinterwordspacing
  
  \bibitem{avm}
  \BIBentryALTinterwordspacing
  Google. {AVM Codec}. Accessed 2022-07. [Online]. Available:
    \url{https://gitlab.com/AOMediaCodec/avm}
  \BIBentrySTDinterwordspacing
  
  \bibitem{Laude_2019}
  T.~Laude, Y.~G. Adhisantoso, J.~Voges, M.~Munderloh, and J.~Ostermann, ``A
    comprehensive video codec comparison,'' \emph{{APSIPA} Transactions on Signal
    and Information Processing}, vol.~8, no.~1, 2019.
  
  \bibitem{Kraenzler2020MMSP}
  M.~{Kr\"anzler}, C.~{Herglotz}, and A.~{Kaup}, ``A comparative analysis of the
    time and energy demand of versatile video coding and high efficiency video
    coding reference decoders,'' in \emph{Proc. IEEE International Workshop on
    Multimedia Signal Processing (MMSP)}, Tampere, Finland, Sep. 2020.
  
  \bibitem{KhernacheBenmoussaBoukhobzaEtAl2021}
  M.~B.~A. Khernache, Y.~Benmoussa, J.~Boukhobza, and D.~Menard, ``{HEVC}
    hardware vs software decoding: An objective energy consumption analysis and
    comparison,'' \emph{Journal of System Architecture}, vol. 115, p. 102004, May
    2021.
  
  \bibitem{Mercat_2021}
  A.~Mercat, A.~Makinen, J.~Sainio, A.~Lemmetti, M.~Viitanen, and J.~Vanne,
    ``Comparative rate-distortion-complexity analysis of {VVC} and {HEVC} video
    codecs,'' \emph{{IEEE} Access}, vol.~9, pp. 67\,813--67\,828, 2021.
  
  \bibitem{Nguyen_2021}
  T.~Nguyen and D.~Marpe, ``Compression efficiency analysis of {AV}1, {VVC}, and
    {HEVC} for random access applications,'' \emph{{APSIPA} Transactions on
    Signal and Information Processing}, vol.~10, no.~1, 2021.
  
  \bibitem{Katsenou_2022}
  A.~Katsenou, J.~Mao, and I.~Mavromatis, ``Energy-rate-quality tradeoffs of
    state-of-the-art video codecs,'' in \emph{Proc. Picture Coding Symposium
    ({PCS})}, San Jose, CA, USA, Dec. 2022.
  
  \bibitem{CWGB005oV1}
  X.~Zhao, Z.~Lei, A.~Norkin, T.~Daede, and A.~Tourapis, ``{AV2} common test
    conditions v1.0,'' document, CWG-B005o v1, Jan. 2021.
  
  \bibitem{HerglotzSpringerReichenbachEtAl2018}
  C.~{Herglotz}, D.~{Springer}, M.~{Reichenbach}, B.~{Stabernack}, and A.~{Kaup},
    ``Modeling the energy consumption of the {HEVC} decoding process,''
    \emph{IEEE Transactions on Circuits and Systems for Video Technology},
    vol.~28, no.~1, pp. 217--229, Jan. 2018.
  
  \bibitem{DavidGorbatovHanebutteEtAl2010}
  H.~David, E.~Gorbatov, U.~R. Hanebutte, R.~Khanna, and C.~Le, ``{RAPL}: Memory
    power estimation and capping,'' in \emph{Proc. ACM/IEEE International
    Symposium on Low-Power Electronics and Design (ISLPED)}, Austin, TX, USA,
    Aug. 2010.
  
  \bibitem{JVET-P0084}
  M.~{Kr\"anzler}, C.~{Herglotz}, and A.~{Kaup}, ``Decoding energy assessment of
    {VTM-6.0},'' Geneva, Switzerland, {document, JVET-P0084}, Oct. 2019.
  
  \bibitem{VCEG-M33}
  G.~Bj{\o}ntegaard, ``Calculation of average {PSNR} differences between {RD}
    curves,'' Austin, TX, USA, {document, VCEG-M33}, Jan. 2001.
  
    \bibitem{Herglotz2023}
    C.~Herglotz, H.~Och, A.~Meyer, G.~Ramasubbu, L.~Eichermüller, M.~Kränzler,
      F.~Brand, K.~Fischer, D.~T. Nguyen, A.~Regensky, and A.~Kaup, ``The
      bj{\o}ntegaard bible -- why your way of comparing video codecs may be
      wrong,'' \emph{IEEE Transactions on Image Processing}, vol.~33, pp.
      987--1001, Jan. 2024.
  
  \bibitem{Herglotz2023a}
  C.~Herglotz, R.~Mons, M.~Kr\"anzler, and A.~Regensky. (2023)
    {Bj{\o}ntegaard-Delta: Scripts for BD calculations with different
    interpolators.} https://github.com/FAU-LMS/bjontegaard. Accessed 2023-11.
  
  \bibitem{WorkingPractices}
  J.~{Str\"om}, K.~{Andersson}, R.~{Sj\"oberg}, A.~{Segall}, F.~{Bossen},
    G.~{Sullivan}, J.-R. {Ohm}, and A.~{Tourapis}, ``Working practices using
    objective metrics for evaluation of video coding efficiency experiments,''
    ITU-T and ISO/IEC, JTC 1, {document, ISO/IEC DTR 23002-8}, Jul. 2020.
  
  \end{thebibliography}

\end{document}